\documentclass[sigconf]{acmart}

\usepackage{booktabs} 
\usepackage{times}
\usepackage{latexsym}
\usepackage{booktabs}
\usepackage{amssymb}
\usepackage{dsfont}
\usepackage{caption}
\usepackage{multirow}
\usepackage{subfig}
\usepackage{amsmath}
\usepackage{todonotes}
\usepackage{multirow}
\usepackage{subfig}
\usepackage{csquotes}
\usepackage{hyperref}
\usepackage{url}
\usepackage{graphicx,wrapfig,lipsum}

\usepackage{arydshln}
\usepackage[markup=underlined]{changes}

\usepackage{xcolor,colortbl}
\usepackage{array}
\newcolumntype{L}[1]{>{\raggedright\let\newline\\\arraybackslash\hspace{0pt}}m{#1}}
\newcolumntype{C}[1]{>{\centering\let\newline\\\arraybackslash\hspace{0pt}}m{#1}}
\newcolumntype{R}[1]{>{\raggedleft\let\newline\\\arraybackslash\hspace{0pt}}m{#1}}

\setcopyright{none}

\copyrightyear{2017}
\acmYear{2017}
\setcopyright{rightsretained}
\acmConference{SIGIR 2017 Workshop on Neural Information Retrieval (Neu-IR'17)}{}{August 7--11, 2017, Shinjuku, Tokyo, Japan} \acmPrice{}
\begin{document}
\title{Toward Incorporation of Relevant Documents in word2vec}
\subtitle{}

\author{Navid Rekabsaz}
\authornote{Funded by: Self-Optimizer (FFG 852624) in the EUROSTARS programme, funded by EUREKA, BMWFW and European Union, and ADMIRE (P 25905-N23) by FWF. Thanks to Joni Sayeler and Linus Wretblad for their contributions in SelfOptimizer.}
\affiliation{%
  \institution{Information \& Software Eng. Group}
  \city{TU Wien} 
}
\email{rekabsaz@ifs.tuwien.ac.at}

\author{Bhaskar Mitra}
\authornote{The author is a part-time PhD student at University College London.}
\affiliation{%
  \institution{Microsoft, UCL\\Cambridge, UK}
}
\email{bmitra@microsoft.com}

\author{Mihai Lupu, Allan Hanbury}
\affiliation{%
  \institution{Information \& Software Eng. Group}
  \city{TU Wien} 
}
\email{lupu/hanbury@ifs.tuwien.ac.at}

\begin{abstract}
Recent advances in neural word embedding provide significant benefit to various information retrieval tasks. However as shown by recent studies, adapting the embedding models for the needs of IR tasks can bring considerable further improvements. The embedding models in general define the term relatedness by exploiting the terms' co-occurrences in short-window contexts. An alternative (and well-studied) approach in IR for related terms to a query is using local information i.e. a set of top-retrieved documents. In view of these two methods of term relatedness, in this work, we report our study on incorporating the local information of the query in the word embeddings. One main challenge in this direction is that the dense vectors of word embeddings and their estimation of term-to-term relatedness remain difficult to interpret and hard to analyze. As an alternative, explicit word representations propose vectors whose dimensions are easily interpretable, and recent methods show competitive performance to the dense vectors. We introduce a neural-based explicit representation, rooted in the conceptual ideas of the word2vec Skip-Gram model. The method provides interpretable explicit vectors while keeping the effectiveness of the Skip-Gram model. The evaluation of various explicit representations on word association collections shows that the newly proposed method outperforms the state-of-the-art explicit representations when tasked with ranking highly similar terms. Based on the introduced explicit representation, we discuss our approaches on integrating local documents in globally-trained embedding models and discuss the preliminary results.

\end{abstract}

%
%



\maketitle

\section{Introduction}
\label{sec:introduction}
Word embedding i.e. representation of words in a low-dimensional space was shown to benefit IR tasks like document retrieval~\cite{rekabsaz2016generalizing,zuccon2015integrating,rekabsaz2017exploration,mitra2016dual}, query expansion~\cite{diaz2016query,zamani2016embedding}, or sentiment analysis~\cite{rekabsaz2017volatility}. 
Despite the relative improvement of various IR tasks, recently Rekabsaz et al.~\cite{rekabsaz2017word} points out the issue of topic shifting when directly using window-based word embedding models (i.e. word2vec~\cite{mikolov2013efficient} and GloVe~\cite{pennington2014glove}). Their study shows a high potential in further performance improvement in document retrieval when filtering out the terms that cause the topic shifting.

One fundamental issue in using word embedding (or most other methods of term similarity) is rooted in the assumption of independence of query terms i.e. the similar terms to a query term are independent of the other query terms. Diaz et al.~\cite{diaz2016query} approach this issue by training separate embedding models on local information of the query i.e. a set of top retrieved documents (top 1000 in their study). They show that the locally-trained word embeddings outperform the global embedding model as the similar terms are relevant to the whole the query. However, as mentioned in the study, it is not a feasible approach because a new model has to be trained at query time (albeit on a small set of documents). The use of local information for retrieval has extensively studied~\cite{lavrenko2001relevance,zhai2001model,lv2009comparative}. Recent studies use the local information to train a relevant-based embedding~\cite{zamani2017relevance} or multi-layer neural network ranking model~\cite{dehghani2017neural}. In contrast, we focus on exploiting window-based term similarities adjusted with local information. 

In this report, we present our ideas and the intermediary steps in this direction, asking \emph{How can we efficiently incorporate the local information in the word embedding models?} The word embeddings, while easy to construct based on raw unannotated corpora, are hardly interpretable, especially in comparison to lexical semantics. It remains opaque what the dimensions of the vectors refer to, or which factors in language (corpus) cause the relatedness of two words. Having interpretable vectors would enable the integration of important topics of the local documents and make a causal analysis possible.

A natural solution to this problem is using explicit representations of words i.e. vectors with clearly-defined dimensions, which can be words, windows of words, or documents. These vectors appeared decades ago as the initial form of word/document representations. In IR, the word-document matrix, populated with an IR weighting schema (e.g. TFIDF, BM25) has been extensively used, at least as a conceptual model. This matrix when being subjected to SVD matrix factorization produces the LSI model~\cite{deerwester1990indexing} with dense word/document representation. The well-known Pointwise Mutual Information (PMI) method is an alternative representation method, rooted in information theory, which provides a word-context matrix based on the co-occurrences between words in contexts.

Despite high interpretability, the explicit representations are too large to be stored in memory (i.e. are not efficient) and have been often shown to be less effective in comparison to the low-dimensional dense vectors. In practice, the memory issue can be mitigated by suitable data structures if the  vectors are highly sparse. Regarding the effectiveness, Levy et al.~\cite{levy2015improving} show competitive performance of a proposed explicit representation (Section~\ref{sec:background}) in comparison to the state-of-the-art word embeddings on a set of word association test collections.
 
As an alternative approach to improve interpretability, some recent reports~\cite{Faruqui2015sparse,sun2016sparse} propose methods to increase the sparsity of the dense vectors. The rationale of these approaches is that by having more sparse vectors, it becomes more clear which dimension of the vectors might be referring to which concepts in language. 

In contrast to this approach, our main contribution is in line with previous studies~\cite{levy2014neural,levy2015improving} on providing fully interpretable vectors by proposing a novel explicit representation, based on the advancements in the neural word embedding methods. Our approach originates from the word2vec Skip-Gram model i.e. the proposed representation benefits from interpretability and also the substantial subtleties of the Skip-Gram model (Section~\ref{sec:method}).

We evaluate our method on the task of retrieving highly similar words to a given word, as an essential scenario in many IR tasks. We show that our proposed explicit representation, as similar to the Skip-Gram model, outperforms the state-of-the-art explicit vectors in selecting highly similar words (Section~\ref{sec:evaluation}).

Given the explicit representation, we discuss our ideas and the preliminary results on integrating the local information of the queries in the representations (Section~\ref{sec:local}). 




\section{Background}
\label{sec:background}
In this section, we review the word2vec Skip-Gram ($SG$) model as well as the state-of-the-art explicit representations.

\subsection{Embedding with Negative Sampling}
The $SG$ model starts with randomly initializing two sets of vectors: word ($V$) and context ($\widetilde{V}$) vectors, both of size $\left|W\right| \times d$, where $W$ is the set of  words in the collection and $d$ is the embedding dimensionality. 

The objective of $SG$ is to find a set of $V$ and $\widetilde{V}$---as the parameters of an optimization algorithm---by increasing the conditional probability of observing a word $c$ given another word $w$ when they co-occur in a window, and decreasing it when they do not. 
In theory, this probability is defined as follows:

\begin{equation}
p(c|w)=\frac{\exp(V_w\widetilde{V}_c)}{\sum_{c' \in W}{\exp(V_w\widetilde{V}_{c'})}}
\label{eq:p_w_c}
\end{equation}
where $V_w$ and $\widetilde{V}_c$ are the word vector of the word $w$ and the context vector of the word $c$, respectively.

Obviously, calculating the denominator of Eq.~\ref{eq:p_w_c} is highly expensive and a bottleneck for scalability. One proposed approach for this problem is \emph{Noisy Contrastive Estimation (NCE)}~\cite{mnih2012fast} method. The NCE method, instead of computing the probability in Eq.~\ref{eq:p_w_c}, measures the probability which contrasts the \emph{genuine} distribution of the words-context pairs (given from the corpus) from a \emph{noisy} distribution. The noisy distribution $\mathcal{N}$ is defined based on the unigram distribution of the words in the corpus. Formally, it defines a binary variable $y$, showing whether a given pair belongs to the genuine distribution: $p(y=1|w,c)$. Further on, Mikolov et al.~\cite{mikolov2013efficient} proposed the \emph{Negative Sampling} method by some simplifications in calculating $p(y=1|w,c)$ (further details are explained in the original papers), resulting in the following formula:

\begin{equation}
p(y=1|w,c)=\frac{\exp(V_w\widetilde{V}_c)}{\exp(V_w\widetilde{V}_c)+1}=\sigma(V_w\widetilde{V}_c)
\label{eq:y1_wc_w2v}
\end{equation}
where $\sigma$ is the sigmoid function ($\sigma(x)=1/(1+\exp(-x))$). Based on this probability, the cost function of the $SG$ method is defined as follows:

\begin{equation}
J=-\sum_{\langle w,c \rangle \in X}{\left[ \log p(y=1|w,c)+k\mathop{{}\mathbb{E}}_{\check{c}_i\sim \mathcal{N}}{\log p(y=0|w,\check{c}_i)}  \right]}
\label{eq:j1}
\end{equation}
where $X$ is the collection of co-occurrence pairs in the corpus, $\check{c}_i$ is each of the $k$ sampled words from the noisy distribution $\mathcal{N}$, and $\mathop{{}\mathbb{E}}$ denotes expectation value, calculated as the average for the $k$ sampled words in every iteration.

In addition, two preprocessing steps dampen the dominating effect of very frequent words: First is \emph{subsampling} which randomly removes an occurrence of word $w$ in the corpus $C$ when the word's frequency in the corpus $f(w,C)$ is more than some threshold $t$ with a probability value of $1-\sqrt{t/f(w,C)}$.  The second is \emph{context distribution smoothing} (\texttt{cds}) which dampens the values of the probability distribution $\mathcal{N}$ by raising them to power $\alpha$ where $\alpha<1$.   

\subsection{Explicit Representation}
A well-known explicit representations is defined based on PMI. In this representation, for the word $w$, the value of the corresponding dimension to the context word $c$ is defined as follows:

\begin{equation}
PMI(w,c)=\log\frac{p(w,c)}{p(w)p(c)}
\label{eq:pmi}
\end{equation}
where $p(w,c)$ is the probability of appearance of $\langle w,c \rangle$ in the co-occurrence collection: $f(\langle w,c \rangle, X)/|X|$ and $p(w)$ is calculated by counting the appearance of $w$ with any other word: $f(\langle w,.\rangle, X)/|X|$ (same for $p(c)$). 

A widely-used alternative is \emph{Positive PMI (PPMI)} which replaces the negative values with zero: $PPMI(w,c)=max(PMI(w,c), 0)$.

Levy and Goldberg~\cite{levy2014neural} show an interesting relation between $PMI$ and $SG$ representations, i.e. when the dimension of the vectors is very high (as in explicit representations), the optimal solution of $SG$ objective function (Eq.~\ref{eq:j1}) is equal to $PMI$ shifted by $\log k$. They call this representation \emph{Shifted Positive PMI (SPPMI)}:

\begin{equation}
SPPMI(w,c)=max(PMI(w,c)-\log(k), 0)
\label{eq:sppmi}
\end{equation}
They further integrate the ideas of subsampling and \texttt{cds} into $SPPMI$. Subsampling is applied when creating the $X$ set by randomly removing very frequent words. The \texttt{cds} method adds a smoothing on the probability of the context word, as follows:

\begin{equation}
PMI_{\alpha}(w,c)=\log\frac{p(w,c)}{p(w)p_{\alpha}(c)} \quad  p_{\alpha}(c)=\frac{f(\langle w,. \rangle, X)^{\alpha}}{\sum_{w'\in W}{f (\langle w',. \rangle, X)}^{\alpha}}
\label{eq:pmialpha}
\end{equation}

They finally show the competitive performance of the $SPPMI$ model on word association tasks to the $SG$ model. Their definitions of $PPMI$ and $SPPMI$ are the current state-of-the-art in explicit representations, against which we will compare our method.

\section{Explicit Skip-Gram}
\label{sec:method}
\begin{table*}[t]
\caption{Word association evaluation. Best performing among explicit/all embeddings are shown with bold/underline.} 
\begin{tabular}{l | c | c c c c c c}
Method & Sparsity& WS Sim. & WS Rel. & MEN & Rare & SCWS & SimLex \\\hline
PPMI & 98.6\% & .681 & .603 & .702 & .309  & .601 & .284\\
SPPMI & 99.6\% & \textbf{.722} & \underline{\textbf{.661}} & .704 & .394 & .571 & \textbf{.296}\\
ExpSG & 0\% & .596 & .404 & .645 & .378 & .549 & .231\\
RExpSG & 0\% & .527 & .388 & .606 & .311 & .507 & .215\\
PRExpSG & 94.1\%& .697 & .626 & \textbf{.711} & \textbf{.406} & \textbf{.614} &.272 \\\hline
SG & 0\% & \underline{.770} & .620 & \underline{.750} & \underline{.488} & \underline{.648}  &  \underline{.367} \\
\end{tabular}
\label{tbl:effectiveness} 
\end{table*}
Let us revisit the $p(y=1|w,c)$ probability in the Negative Sampling method (Eq.~\ref{eq:y1_wc_w2v}) i.e. the probability that the co-occurrence of two terms comes from the training corpus and not from the random distribution. The purpose of this probability in fact shares a meaningful relation with the conceptual goal of the $PMI$-based representations i.e. to distinguish a genuine from a random co-occurrence. Based on this idea, an immediate way of defining an explicit representation would be to use  Eq.~\ref{eq:y1_wc_w2v} as follows:

\begin{equation}
ExpSG(w,c)=p(y=1|w,c)=\sigma(V_w\tilde{V}_c)
\label{eq:expsg}
\end{equation}

This \emph{Explicit Skip-Gram (ExpSG)} representation assigns a value between 0 to 1 to the relation between each pair of words. It is however intuitive to consider that the very low values do not represent a meaningful relation and can potentially introduce noise in computation. Such very low values can be seen in the relation of a word to very frequent or completely unrelated words. We can extend this idea to all the values of $ExpSG$, i.e. some portion (or all) of every relation contains noise. 

%

To measure the noise in $ExpSG(w,c)$, we use the definition of noise probabilities in the Negative Sampling approach: the expectation value of $p(y=1|w,c)$  where $c$ (or $w$) is randomly sampled from the dictionary for several times. Based on this idea, we define the \emph{Reduced Explicit Skip-Gram (RExpSG)} model by subtracting the two expectation values from $ExpSG$:

\begin{equation}
RExpSG(w,c)=ExpSG(w,c)-\mathop{{}\mathbb{E}}_{\check{c}\sim \mathcal{N}}{p(y=1|w,\check{c})} -\mathop{{}\mathbb{E}}_{\check{w}\sim \mathcal{N}}{p(y=1|\check{w},c)} 
\label{eq:rexpsg}
\end{equation}
Since the expectation values can be calculated off-line, in contrast to Negative Sampling, we compute it over all the vocabulary:

\begin{equation}
\mathop{{}\mathbb{E}}_{\check{w}\sim \mathcal{N}}{p(y=1|\check{w},c)}=\frac{\sum_{i=1}^{|W|}{f(\check{w}_i, C) \cdot \sigma(V_{\check{w}_i}\widetilde{V}_c)}}{\sum_{i=1}^{|W|}{f(\check{w}_i, C) }}
\label{eq:expectation2}
\end{equation}
For the sampling of	the context word $\check{c}$, similar to $SG$ and $PMI_{\alpha}$, we apply \texttt{cds} by raising frequency to the power of $\alpha$, as follows:
\begin{equation}
\mathop{{}\mathbb{E}}_{\check{c}\sim \mathcal{N}}{p(y=1|w,\check{c})}=\frac{\sum_{i=1}^{|W|}{f(\check{c}_i, C)^{\alpha} \cdot \sigma(V_w\widetilde{V}_{\check{c}_i})}}{\sum_{i=1}^{|W|}{f(\check{c}_i, C)^{\alpha} }} 
\label{eq:expectation1}
\end{equation}

Similar to $PPMI$, our last proposed representation removes the negative values. The \emph{Positive Reduced Explicit SkipGram (PRExpSG)} is defined as follows:
\begin{equation}
PRExpSG(w,c)=max(RExpSG(w,c), 0)
\label{eq:prexpsg}
\end{equation}

Setting the values to zero in $PRExpSG$ facilitates the use of efficient data structures i.e. sparse vectors. We analyze the efficiency and effectiveness of the explicit representations in the next section.

\section{Evaluation of Explicit Skip-Gram}
\label{sec:evaluation}
To analyze the representations, we create a Skip-Gram model with 300 dimensions on the Wikipedia dump file for August 2015 using the gensim toolkit~\cite{rehurek_lrec}. As suggested by Levy et al.~\cite{levy2015improving}, we use a window of 5 words, negative sampling of $k=10$, down sampling of $t=10^{-5}$, a \texttt{cds} value of $\alpha=0.75$, trained on 20 epochs, and filtering out  words with frequency less than 100. The final model contains 199851 words. The same values are used for the common parameters in the $PPMI$ and $SPPMI$ representations.

We conduct our experiments on 6 word association benchmark collections. Each collection contains a set of word pairs where the association between each pair is assessed by human annotators (annotation score). The evaluation is done by calculating the Spearman correlation between the list of pairs scored by similarity values versus by human annotations. The collections used are: WordSim353 partitioned into Similarity and Relatedness~\cite{agirre2009study}; MEN dataset~\cite{bruni2014multimodal}; Rare Words dataset~\cite{luong2013better}; SCWS~\cite{HuangEtAl2012}; and SimLex dataset~\cite{hill2016simlex}.


The evaluation results for the explicit representations as well as $SG$ are reported in Table~\ref{tbl:effectiveness}. The bold values show the best performing explicit representation and the values with underline refer to the best results among all representations. Based on the results, $PRExpSG$ and $SPPMI$ show very similar performance (in 3 benchmarks $PRExpSG$ and in the other 3 $SSPMI$ shows the best performance), both considerably outperforming the other explicit representations. As also shown in previous studies~\cite{levy2015improving}, $SG$ in general performs better than the best performing explicit representations. 

However, when considering downstream tasks, despite the pervasive use of word association benchmarks, they do not provide a comprehensive assessment on many subtleties of representations which might be crucial (see Faruqui et al.~\cite{faruqui2016problems}). For example, in tasks such as query expansion~\cite{diaz2016query,zamani2016embedding} or the integration of embeddings in IR models~\cite{rekabsaz2016generalizing,rekabsaz2017exploration}, what is expected from an effective word representation is a set of \emph{highly} similar words for each query word. 

To have a more relevant evaluation for these sort of tasks, we need to consider (1) the word \emph{similarity} benchmarks (in contrast to \emph{relatedness}) and (2) the effectiveness of the representation on highly similar words. Among the benchmarks, SimLex is a recent collection, specifically designed to evaluate the concept of word similarity. In particular, SimLex's creators criticize the WordSim353 Similarity collection as it does not distinguish between word similarity and relatedness~\cite{hill2016simlex}. We therefore focus on the SimLex collection for further evaluations. To assess the effectiveness of representations on highly similar pairs, we extract subsets of the SimLex collection that have higher annotation score than a threshold, and calculate the Spearman correlation of the results, separately for each subset. We conduct evaluation on 10 subsets with the thresholds from 9 (highly similar) to 0 (all pairs).

\begin{figure}[t]
 \includegraphics[width=0.34\textwidth]{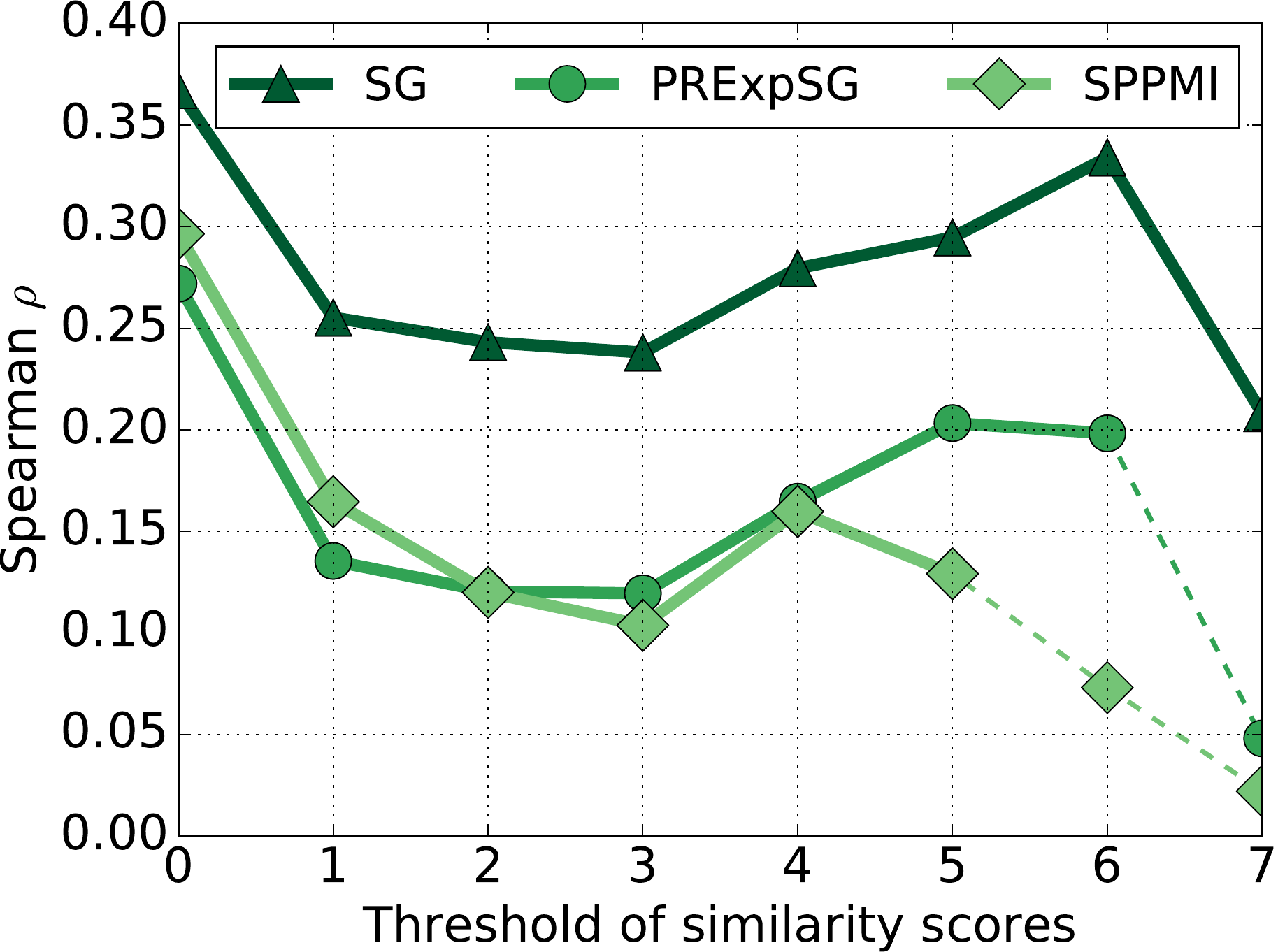}
\caption{Correlation of subsets of SimLex for various similarity thresholds.}
 \label{fig:w2wsim_simlex}
\end{figure}

Figure~\ref{fig:w2wsim_simlex} shows the evaluation results on the subsets for the $SG$, $PRExpSG$, and $SPPMI$ representations. When the  threshold is higher than 7, none of the models has significant correlation values ($p<0.05$) and are therefore not depicted. The non-significant results for $PRExpSG$ and $SPPMI$ are indicated with dashed lines. The $SG$ model constantly shows high correlation values over all the thresholds. Interestingly, while the $PRExpSG$ has slightly worse performance than $SPPMI$ at lower thresholds, it reaches better correlation results at high thresholds. We argue that the better performance for highly similar pairs is due to the conceptual similarity of $PRExpSG$ to the $SG$ model i.e. $PRExpSG$ benefits from the subtleties of the $SG$ model.

Finally, let us have a look at the sparsity of the explicit representations, reported in Table~\ref{tbl:effectiveness}. The $PRExpSG$ and $SPPMI$ representations are highly sparse, making them amenable to storage in volatile memory in practical scenarios. It is also worth noticing that in contrast to $SPPMI$, in the $PRExpSG$ vectors, there might be non-zero values also for pairs of terms that do not necessarily ever co-occur in text. This characteristic---inherited from the $SG$ model---is arguable the reason for better performance in high-similarity values, by mitigating the sparseness problem of a corpus.


\section{Integration of Local Information}
\label{sec:local}
In this section, we discuss our ideas on incorporating the information of the local documents of a query in an explicit representation. Each cell value of any of the discussed explicit representations defines the degree of a genuine co-occurrence relationship between the term $w$ to a context term $c$. We refer to the relation as $v(w,c)$. Our objective is to alter $v$ to $\hat{v}$ based on the information in the local document i.e. the top $k$ retrieved documents ($k$ has a small number like 10). 

We define the relation between $v$ and $\hat{v}$ by a logistic function as follow:

\begin{equation}
\hat{v}(w,c)=\frac{1}{1+e^{-(a+b \cdot f(w,c,F))}} v(w,c)
\label{eq:delta1}
\end{equation}
where $a$ and $b$ are model parameters, and $f$ is a function to incorporate the query's local information. In the following, we suggest various definition for $f$. 

The first suggestion only keeps the dimensions (context terms) of the representation whose corresponding terms appear in the relevant documents and sets the values related to other dimensions to zero.
\begin{equation}
f_1(w,c,F)=f_1 (c,F)=\mathds{1}\big[ f(c,F) > 0 \big] 
\label{eq:delta1}
\end{equation}
where $\mathds{1}$ is the indicator function, and $F$ is the collection of local documents. As mentioned in the Eq.~\ref{eq:delta1}, it is only based on the context terms $c$ and independent of $w$ (it also holds in some other formulas in the following as it is mentioned in the equation). Our initial results using the $f_1$ method show little differences in the orders of top similar terms to the query terms when comparing the global embedding with the locally-adapted representation. We argue that it is due to the neglect of the weights of the terms in the local documents. In fact, since (very probably) there are still irrelevant documents in the top retrieved ones, by considering only the occurrence of the terms some irrelevant dimensions/terms still affect the similarity calculation.

To exploit the importance of the terms of the local documents, in the second method, we use the probability of the occurrence of the context term $c$ in the global and local documents, defined as follows: 
\begin{equation}
f_2 (w,c,F) = f_2 (c,F) = \frac{p(c|F)}{p(c|C)} = \frac{f(c,F) / \sum_{d \in F}{|d|}} {f(c,C) / \sum_{d \in C}{|d|}} 
\label{eq:delta2}
\end{equation}

The third method reflects the idea of the $PMI$-based representations by measuring the probability of co-occurrences of $w$ and $c$ in some defined contexts in local and global documents.
\begin{equation}
f_3 (w,c,F) = \frac{p(w,c|X_F)}{p(w,c|X_C)} = \frac{f(\langle w,c\rangle,X_F) / |X_F|}{f(\langle w,c \rangle,X_C) / |X_C|} 
\label{eq:delta3}
\end{equation}
where $X_F$ and $X_C$ refer to the local and global co-occurrence sets respectively. The context in $f_3$ can be defined as short-window around $w$, in paragraph-level, or whole the document.

The next method exploits the idea of the relevance model~\cite{lavrenko2001relevance} and is defined as follows:
\begin{equation}
f_4 (w,c,F) = f_4 (c) = p(c|\Theta_F) = \sum_{\theta_d \in \Theta_F}{p(c|\theta_d) \prod_{q \in Q}{p(q|\theta_d)}}
\label{eq:delta4}
\end{equation}
where $Q$ is the set of query terms $q$, $\Theta_F$ is the collection of documents' language models $\theta_d$ for each document $d \in F$. The document language models is calculated using with Dirichlet smoothing with $\mu=1500$.


The last approach expands the $f_4$ method by also using the probability of $w$ in the joint probability $p(w,c|\theta_d)$. We assume the independence of $w$ and $c$ when conditioned on $\theta_d$ i.e. $p(w,c|\theta_d)=p(w|\theta_d)p(c|\theta_d)$, resulting to the following definition:
\begin{equation}
f_5 (w,c,F) = p(w,c|\Theta_F) = \sum_{\theta_d \in \Theta_F}{p(w|\theta_d)p(c|\theta_d) \prod_{i \in Q}{p(q|\theta_d)}}
\label{eq:delta5}
\end{equation}

We expect that the report provides the ground for circulating discussions on the theme of the study and suggested ideas.

\section{Conclusion}
\label{sec:conclusion}
Incorporating the important topics of the local documents in word embeddings is a crucial step for specializing the embedding models for IR tasks. To move toward this direction, we propose a novel explicit representation of words by capturing the probability of genuine co-occurrence of the words, achieved from the word2vec Skip-Gram model. The proposed representation inherits the characteristics of the Skip-Gram model while making it possible to interpret the vector representations. The evaluation on term association benchmarks shows similar results to the state-of-the-art explicit representations, but our method outperforms the state-of-the-art in the scenario of retrieving top-similar words to a given word. Further on, based on the introduced explicit representation, we discuss our proposed methods to redefine the terms' vectors using the importance of the terms in the top-retrieved documents.  

\bibliographystyle{ACM-Reference-Format}
\bibliography{sigproc} 

\end{document}